\documentclass[10pt,aps,prb,twocolumn,noshowpacs]{revtex4-1}
\usepackage{amsmath,graphicx,amsbsy,amssymb,amsmath,epsfig,latexsym,wasysym,pifont,mathrsfs,natbib}
\usepackage{float} \newfloat{widefig}{thp}{lop} \usepackage{comment}

\usepackage{array, makecell} \usepackage{verbatim} \usepackage{datetime}
\usepackage{multirow,array} \usepackage{hyperref} \usepackage{color} \usepackage{setspace}
\usepackage{gensymb}
\usepackage{subcaption}
\usepackage{graphicx}
\usepackage{lipsum}
\usepackage[export]{adjustbox}

\hypersetup{colorlinks,%,%
  linkcolor=blue,%
  citecolor=blue,%
  urlcolor=blue}

\begin{document}
\title{\textbf{Tunable magnetism in Nitride MXenes:consequences of atomic layer stacking}}

\author{Himangshu Sekhar Sarmah}
\email[]{shimangshu@iitg.ac.in}
\affiliation{Department of Physics, Indian Institute of Technology
  Guwahati, Guwahati-781039, Assam, India.}    
\author{Subhradip Ghosh}
\email{subhra@iitg.ac.in} \affiliation{Department of Physics,
  Indian Institute of Technology Guwahati, Guwahati-781039, Assam,
  India.} 

\begin{abstract}
We have performed Density Functional Theory (DFT) based calculations to investigate the effects of stacking patterns on the electronic and magnetic properties of several Nitride MXenes. MXenes, a relatively new addition to the family of two-dimensional materials, have exhibited fascinating properties on several occasions, primarily due to their compositional flexibility. However, compared to Carbide MXenes, Nitride MXenes are much less explored. Moreover, the structural aspects of MXenes and the tunability it may offer have not been explored until recently. In this work, we have combined these two less-explored aspects to examine the structure-property relations in the field of magnetism. We find that in the family of M$_{2}$NT$_{2}$ (M=Sc, Ti, V, Cr, Mn; T=O, F) MXenes, the stacking of transition metal planes has a substantial effect on the ground state and finite temperature magnetic properties. We also find that the electronic ground states can be tuned by changing the stacking pattern in these compounds, making the materials appropriate for applications as magnetic devices. Through a detailed analysis, we have connected the unconventional stacking pattern-driven tunability of these compounds with regard to electronic and magnetic properties to the local symmetry, inhomogeneity (or lack of it) of structural parameters, and electronic structures.  
\end{abstract}

\pacs{}

\maketitle

\section{Introduction\label{intro}}
MXenes, discovered barely over a decade ago \cite{naguib2011two} the new addition to the family of two-dimensional (2D) materials, have caught the attention of materials scientists and technologists due to their high electrical conductivity, large surface area, tuneable structure, among many other properties, that can be exploited to obtain superior functional materials. This expectation is fuelled after spectacular successes of MXenes in variety of applications such as supercapacitors\cite{lukatskaya2017ultra,lukatskaya2013cation,fan2018modified}, lithium-ion batteries \cite{naguib2013new,ahmed2016h,du2017environmental}, and photocatalytic water splitting \cite{shuai2023recent,bai2021recent}, to name a few. MXenes, obtained from their three-dimensional counterpart, the MAX compounds, most commonly through chemical exfoliation process \cite{hope2016nmr}, have $(n+1)$ layers of hexagonal close-packed transition metals (M) intercalated by $n$ layers of hexagonal close-packed C or N (X) with a face centered cubic (fcc) stacking, where $n=1-3$. Such stacking of M and X layers is the most commonly known stacking pattern observed in MXenes, known as ABC stacking. The chemical exfoliation from MAX compounds introduces passivation of surfaces by functional groups T$_{x}$, making the chemical formula of MXene M$_{n+1}$X$_{n}$T$_{x}$; the most common T being -O,-F and -OH.

Though almost all MXenes synthesized so far have the layers stacked in ABC way, a very recent experiment on synthesizing V$_{2}$N and Mo$_{2}$N from their respective parent carbides by Ammonia treatment observed that Mo$_{2}$N crystallized in a structure with hcp $D_{3h}$ symmetry instead of the usual trigonal $D_{3d}$ symmetry, seen in MXenes with ABC stacking sequence \cite{urbankowski20172d}. This stacking sequence can be described as ABA. Subsequently, a comprehensive study of 54 different MXenes \cite{gouveia2020mxenes}using robust first-principles Density Functional Theory (DFT) \cite{dft} based calculations was performed to assess the possibility of alternative ABA stacking in MXenes. The study found that the ABA stacking is realized more in the thinnest ($n=1$) nitride MXenes where their surfaces are passivated by -O. The possibility of obtaining the ABA stacking pattern is also dependent on the number of $d$-electrons in the transition metal constituents of the MXenes. This investigation provided enough hints towards the alteration of surface phenomena due to changes in the stacking pattern. That this can have serious consequences on the functional properties of these compounds is proved in -O terminated carbide MXenes where ABA stacking is predicted to infuse greater electrocatalytic activity on their surfaces, boosting the hydrogen evolution reaction \cite{bixene}.

Magnetism is one area where structural aspects have significant influences. Plethora of studies on magnetic MXenes have demonstrated that the composition \cite{khazaei,junjinhe,kumar2017tunable,janusV,imxene,mxene-nanomaterials}, surface functional groups \cite{chen2015,frey2019,zheng2019} and number of layers \cite{wei2022,sarmah1} affect magnetic properties substantially. Changes in atomic arrangements can alter the bonding and magnetic exchange interactions, resulting in changes in the electronic and magnetic ground states and even finite temperature properties like magnetic transition temperatures. All these have serious consequences on the feasibility of using these compounds in practical applications. In what follows, we perform a comprehensive study of several thinnest Nitride MXene M$_{2}$NT$_{2}$ (M=Sc, Ti, Cr, V, Mn; T=O, F) with a focus on a comparative analysis of their electronic and magnetic properties when stacking pattern changes between ABC and ABA. Finite temperature magnetic properties and magneto-crystalline anisotropy energy (MAE), both essential with regard to using the materials as devices, have also been investigated and analyzed from a microscopic point of view. We find that the stacking pattern can be used as another agency to tune the magnetic properties of these MXenes to a great extent.      
\section{Computational Details}
Density Functional Theory (DFT)\cite{dft} realized within the Projector Augmented Wave (PAW) method \citep{kresse1999ultrasoft}, and implemented in the Vienna ab initio Simulation Package (VASP) \citep{kresse1996efficient}, has been used for all calculations. The exchange-correlation part of the Hamiltonian is approximated using Generalized Gradient Approximation (GGA), parameterized by Perdew-Burke-Ernzerhof (PBE)\citep{perdew1996generalized}. DFT+U method\cite{anisimov1997first} has been used to include  strong correlations among electrons of the transition metal constituents. The U parameters for Sc, Ti, V, Cr, and Mn are taken as 3.0 eV, 4.0 eV,3.0 eV, 4.0 eV and 4.0 eV, respectively, after carefully checking the literature\cite{liu2023mxene,kumar2017tunable,karmakar2020first}. To include the Van Der- Waals interactions, the DFT-D3 method was employed\citep{grimme2010consistent}. Plane waves up to 600 eV and a $\Gamma$-centred $8\times 8\times 1 (18 \times 18 \times 1)k$-mesh for geometry optimization (electronic structure calculations) have been used. The convergence criterion for energies (forces) is set to $ 10^{-6} $ eV(0.01 eV/{\AA}). For magnetic anisotropy energy (MAE) calculation, spin-orbit coupling (SOC) and a denser $k$-mesh are used. The MXene mono-layers are simulated for all calculations by considering a large vacuum of 20 {\AA} in the perpendicular direction.

The dynamical stability of the systems considered is verified by computing their phonon dispersion relations by the supercell method as implemented in the PhonoPy package \cite{togo2015first}. $ 4\times 4\times1$ supercells are used to compute the phonons. The convergence criteria for energy is kept at a high value of $ 10^{-8} $ eV to achieve good convergence. The magnetic exchange interactions are calculated using the magnetic force theorem (MFT)\cite{liechtenstein1987local}, implemented in Relativistic Spin Polarised Toolkit RSPt\cite{wills2010full}. The angular momentum cutoff is taken as l\textsubscript{max} = 8, Brillouin zone integration is done by the tetrahedron method, and the convergence criterion is set to $ 10 ^{-8} $ for self-consistence cycles. The Matsubara frequency is set at 1500 after carefully checking the convergences. The exchange parameters obtained are then used to estimate the magnetic ordering temperature through the Classical Monte Carlo simulation (MCS) method as implemented in UppASD code\citep{eriksson2017atomistic}. Calculations are performed for three ensembles in supercell of size $ 30\times 30\times1$.Periodic boundary conditions are taken into account. At each temperature, 80,000 Monte Carlo steps are considered to achieve energy convergence.
\section{Results and Discussions}
\subsection{Structural models and stability of functionalised MXenes}
\begin{figure}
    \includegraphics[height=8cm, width=8.00 cm]{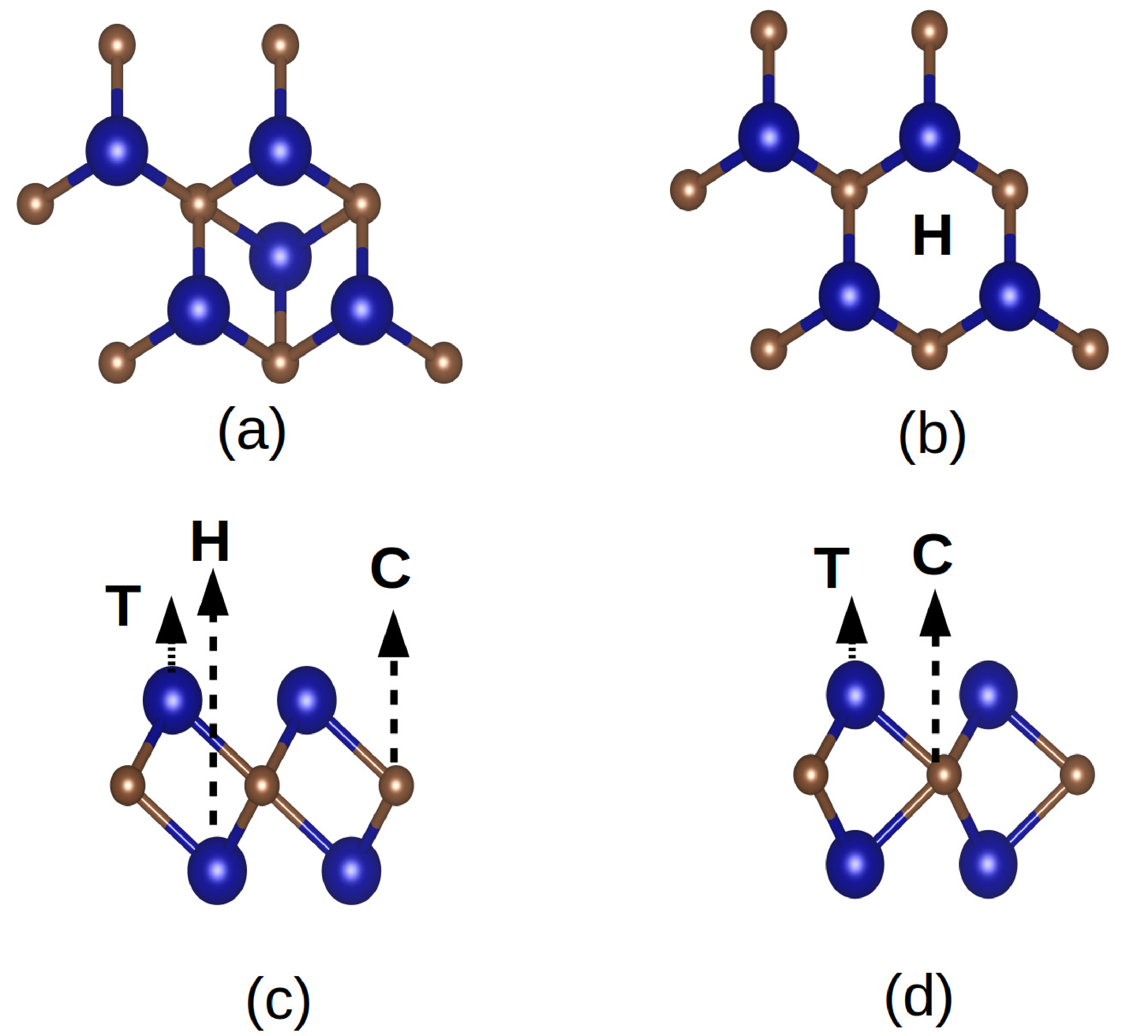}
     \caption{Top (a) ((b)) and side (c)((d))  view of ABC (ABA)  stacked MXene, M\textsubscript{2}X. Blue(brown) spheres are M(X) atoms.}
    \label{Fig:1}
\end{figure}
In Figure \ref{Fig:1}, we present the structure of monolayer MXene. Figures \ref{Fig:1}(a) and (c) ((b) and (d)) are the structures with ABC (ABA) stacking pattern. The difference between the two stacking types is in the alignments of the M layers with respect to each other. In the ABC stacking, the layers are not aligned. A sliding motion of one of the M layers would bring it to the ABA structure. Based upon the sites of functionalization, the structures of M$_{2}$NT$_{2}$ can be different. In the case of usual ABC stacking, it is well known that there are three possible sites of functionalization, H, C, and T (Figure \ref{Fig:1}(c)). Site H is the hollow site associated with the M atom, while site C is the hollow site associated with the X atom. The site T is the one right on top of M. In ABA stacking, C and T sites are identical to those in ABC stacking. The hollow site H now is the one at the center of the hexagon consisting of M and X atoms (Figure \ref{Fig:1} (b)). Consequently, for either stacking, there are four possible structural models of surface functionalization: HH, CC, HC(CH), and TT. In the HH(CC) model, functional groups on both surfaces occupy hollow site H(C), while the site occupancy pattern is mixed for the HC model. Existing results on the MXenes considered here are with ABC stacking where all of them were considered to stabilize in the HH model \cite{kumar2017tunable,sc2no2,mxene-nanomaterials,wei2022,sc2nf2}. Therefore, we consider the 10 MXenes mentioned in Section \ref{intro} in ABA stacking only and calculate the total energy of each one of them in four structural models. In each case, the structural model that produces the lowest total energy is considered the structural model for that particular compound. We calculate its phonon dispersion relations to check whether a given compound is dynamically stable with the optimized structural model. In Figure S1, supplementary information, we show the phonon spectra of all 10 MXenes considered here. We find that except Mn$_{2}$NO$_{2}$ and Ti$_{2}$NO$_{2}$, all are dynamically stable in ABA stacking.The structural models of functionalization for these eight are listed in Table \ref{tab1}. We find that the compounds are evenly distributed among HH and CC models of functionalization. Since Mn$_{2}$NO$_{2}$ and Ti$_{2}$NO$_{2}$ are dynamically stable in ABC stacking only\cite{wang2017first,jena2022surface}, they are not considered for further investigation. 
\begin{table}
\caption{\label{tab1} Structural Models of functionalisation for ABA stacked MXenes considered in this work. }
\begin{tabular}{m{0.30\textwidth}m{0.15\textwidth}}
 \hline
Compounds & Structural Model  \\ 
 \hline
 Cr\textsubscript{2}NF\textsubscript{2},Sc\textsubscript{2}NO\textsubscript{2},Mn\textsubscript{2}NF\textsubscript{2} ,Sc\textsubscript{2}NF\textsubscript{2}& HH \\
 \hline
 Cr\textsubscript{2}NO\textsubscript{2},V\textsubscript{2}NO\textsubscript{2},Ti\textsubscript{2}NF\textsubscript{2},V\textsubscript{2}NF\textsubscript{2},
  & CC \\
 \hline
 \end{tabular}
 \end{table}
 \begin{figure*}
    \includegraphics[height=12cm, width=14.00 cm]{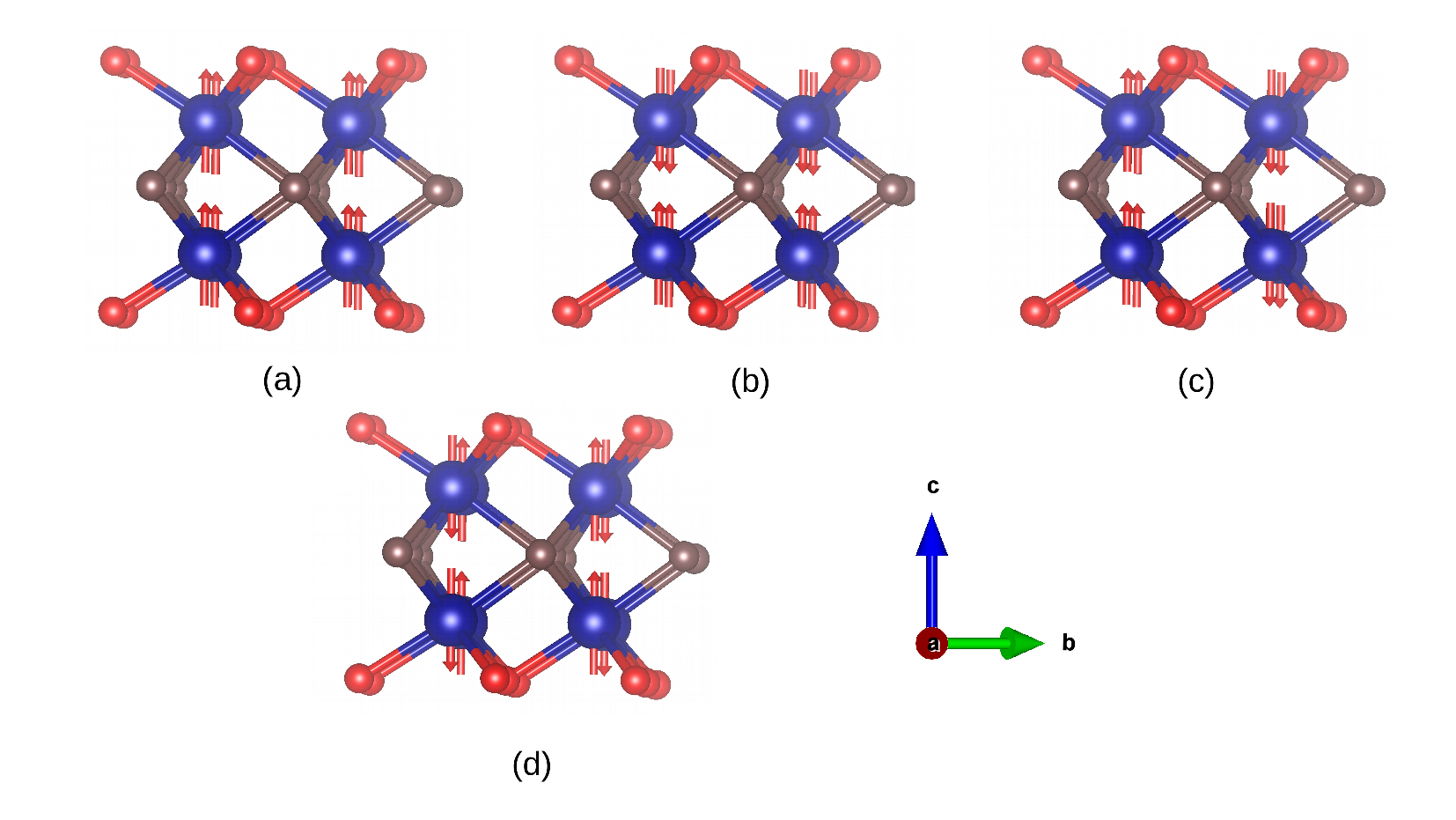}
     \caption{Various magnetic configurations considered in this work to find the magnetic ground state of a given MXene. (a)-(d) stand for FM, AFM1, AFM2 and AFM3 configurations, respectively. Blue, brown and red spheres are  M, N and T in M$_{2}$XT$_{2}$ MXene.}
    \label{Fig:2}
\end{figure*}

\subsection{Stacking dependent magnetic and electronic ground states}
In order to investigate the impact of stacking on the magnetic and electronic ground state properties of the 8 MXenes that are dynamically stable in both stacking patterns, we perform spin-polarised calculations first to determine the magnetic ground states. Four different magnetic configurations, shown in Figure \ref{Fig:2}, are considered for this purpose: 
(a) Ferromagnet(FM) where all spins are aligned along the $c$-axis, (b) AFM1(AFM2) where antiferromagnetic alignment of spins is along $c(b)$ axis only and (c) AFM3 where antiferromagnetic alignment of spins are along both $a$ and $b$ axes. Supercells of dimensions $2\times2\times1$ have been constructed to obtain these four magnetic configurations. The magnetic ground state of a system with a particular stacking is then determined by identifying the magnetic configuration with the lowest total energy. The ground state magnetic configurations and the atomic magnetic moments are presented in Table \ref{tab:Table2}.
\subsubsection{Ground states of MXenes in ABC stacking: comparison to existing results}
\begin{figure*}
    \includegraphics[height=9cm, width=16.00 cm]{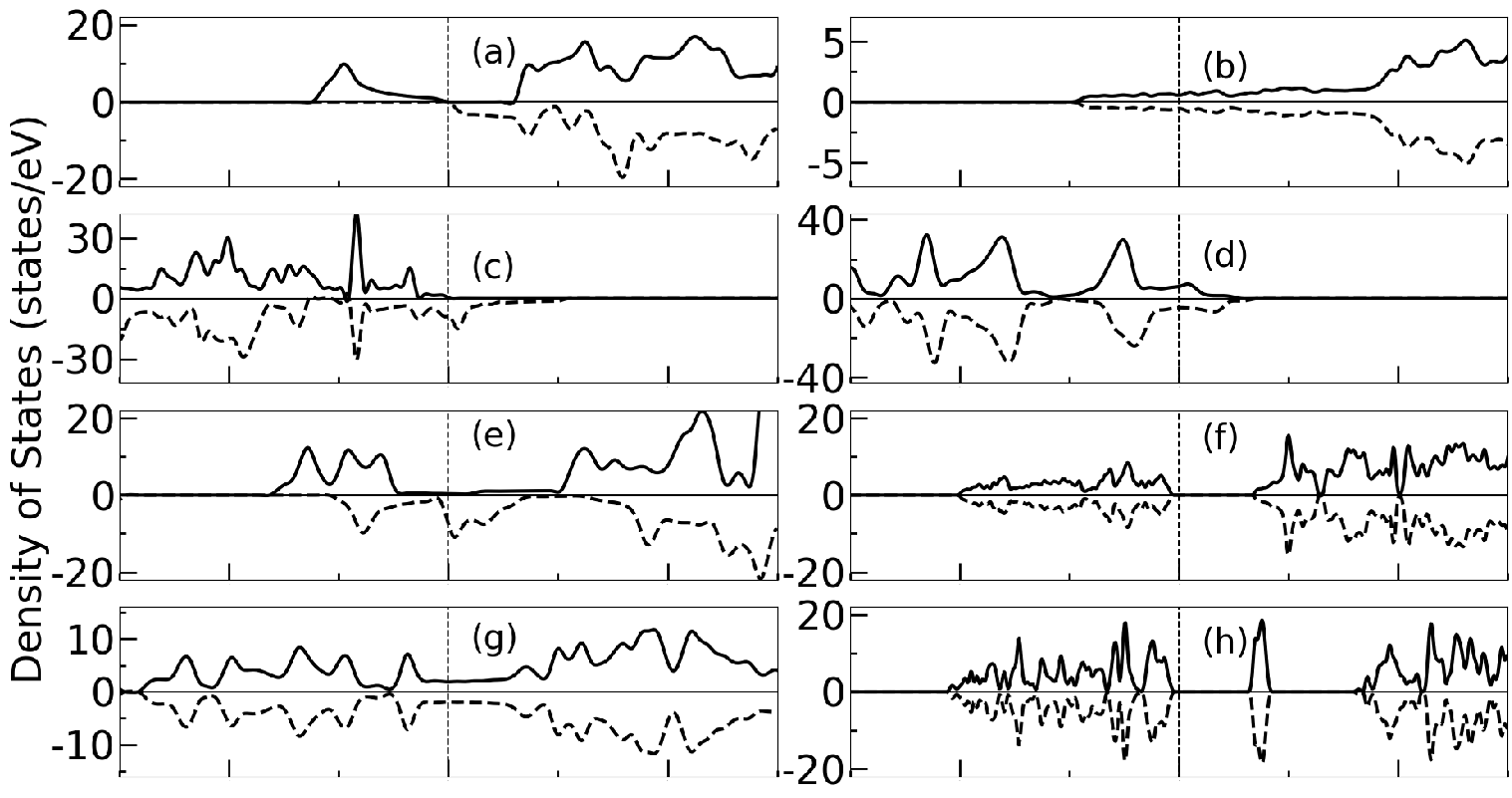}
    \vspace{0.0001cm}  
    \includegraphics[height=9cm, width=16.00 cm]{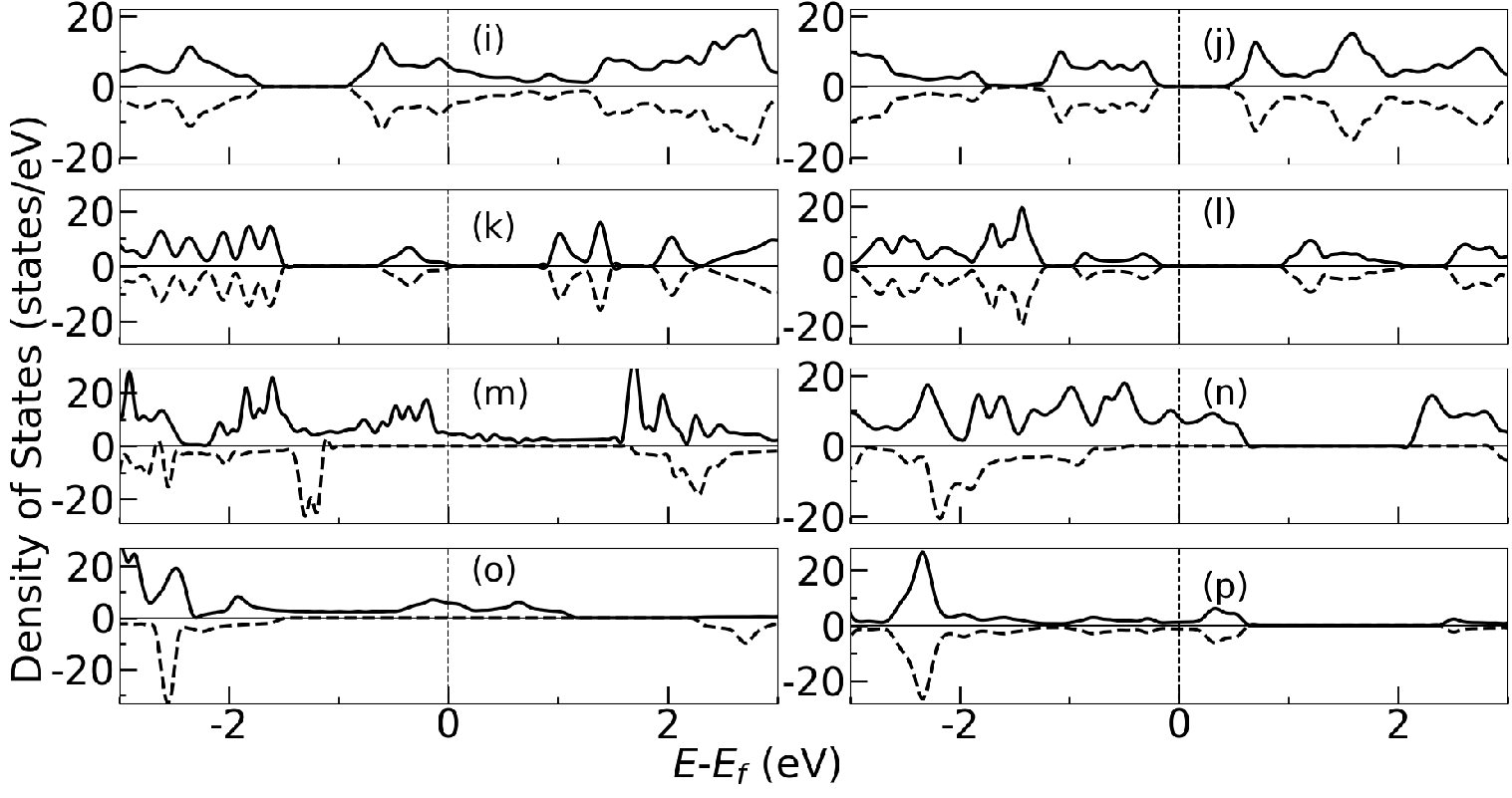} 
     \caption{Total  densities of states of ABA (left panel ) and ABC (right panel) stacked MXenes Sc\textsubscript{2}NF\textsubscript{2}(a),Sc\textsubscript{2}NO\textsubscript{2}(b), Ti\textsubscript{2}NF\textsubscript{2}(c),V\textsubscript{2}NF\textsubscript{2}(d), V\textsubscript{2}NO\textsubscript{2}(e),Cr\textsubscript{2}NF\textsubscript{2}(f),Cr\textsubscript{2}NO\textsubscript{2}(g),Mn\textsubscript{2}NF\textsubscript{2}(h)}
    \label{Fig:3}
\end{figure*}

We first discuss our results with ABC stacking since comparison is possible with other calculations. Comparison with DFT+U calculations from Reference \onlinecite{kumar2017tunable} shows that for Ti$_{2}$NF$_{2}$, V$_{2}$NF$_{2}$ and Cr$_{2}$NO$_{2}$ magnetic ground states from our calculations exactly match with their results. The magnetic ground state of V$_{2}$NO$_{2}$ from our calculations, on the other hand, matches with that from Reference \onlinecite{mxene-nanomaterials}. In case of Cr$_{2}$NF$_{2}$ and Mn$_{2}$NF$_{2}$, the results differ from one another. In both cases, our calculations predict AFM3 to be the ground state. The reason for the difference between our results and that of Reference \onlinecite{kumar2017tunable} is due to the fact that they did not consider AFM3 as one of the possible configurations while determining the ground state. In both cases, we find that the ground state claimed in Reference \onlinecite{kumar2017tunable} is the first one above ground state in our calculations. This implies that had the AFM3 configuration not been considered in our calculations, our results would have agreed with that of Reference \onlinecite{kumar2017tunable} for these two compounds as well. In the case of V$_{2}$NO$_{2}$, the origin of different ground states obtained from our calculations and that of Reference \onlinecite{kumar2017tunable} is their inability to obtain an AFM1 state with unequal atomic moments. On the other hand, the differences between our calculated ground state and that obtained in Reference \onlinecite{mxene-nanomaterials} can be due to the use of different U values in the DFT+U calculations for Cr$_{2}$NF$_{2}$. Moreover, the origin of AFM1 configuration as the ground state of V$_{2}$NF$_{2}$ as reported in Reference \onlinecite{mxene-nanomaterials} is the small energy difference ($\sim$ 4 meV/atom)between their AFM1 and AFM2 configurations. The only available DFT calculation for Sc$_{2}$NF$_{2}$ using the exact exchange HSE06 exchange-correlation functional \cite{sc2nf2} is done on a non-magnetic (NM) state. We, too, get the NM as the ground state in our total energy calculations. For Sc$_{2}$NO$_{2}$, a DFT+HSE06 calculation \cite{sc2no2} obtains an FM ground state while we find the system to be NM. This discrepancy can be an artifact of using an exchange-correlation functional that addresses the localized states of transition metals better than DFT+U. This can be inferred from the values of atomic magnetic moments in Table \ref{tab:Table2}. In Reference \onlinecite{sc2no2}, Sc atoms have no moment, as is the case for our calculations. The FM state obtained in their calculation is due to induced spin polarisation in the N atom. According to their calculations, a total moment of about $\sim 1 \mu_{B}$ is due to N moment only. For the six other materials with different magnetic ground states, atomic magnetic moments obtained in our calculations have an excellent agreement with the existing DFT+U calculations \cite{kumar2017tunable,mxene-nanomaterials}. With regard to electronic ground states, our calculations have substantial agreement with the existing results. The electronic ground states in our calculations can be understood from the average densities of states (Figure \ref{Fig:3} (b),(d),(f),(h),(j),(l),(n),(p)). We find that Ti$_{2}$NF$_{2}$,V$_{2}$NF$_{2}$, V$_{2}$NO$_{2}$ and Cr$_{2}$NF$_{2}$ are semiconductors, in agreement with Reference \onlinecite{mxene-nanomaterials}. Among the remaining four, Sc$_{2}$NF$_{2}$, Sc$_{2}$NO$_{2}$ and Mn$_{2}$NF$_{2}$ are metals while Cr$_{2}$NO$_{2}$ is a half-metal. The electronic ground states of Sc$_{2}$NF$_{2}$ and Cr$_{2}$NO$_{2}$ agree with those obtained in Reference \onlinecite{sc2nf2} and \onlinecite{kumar2017tunable}, respectively. Differences is observed in cases of Sc$_{2}$NO$_{2}$ which is a spin gapless semiconductor and Mn$_{2}$NF$_{2}$, a half metal. The reason for this difference for the former is the usage of HSE06 exchange-correlation functional induced localization in Reference \onlinecite{sc2no2} and consideration of FM magnetic configuration for the latter. Thus, overall, our results have substantial agreement with existing DFT-based calculations as far as magnetic ground states, magnetic moments, and electronic ground states with ABC stacking are concerned.
\subsubsection{Comparison of properties in ABC and ABA stacking}
Having established the quality of our calculations, we now investigate the effects due to changes in stacking patterns. We make a comparative analysis of properties obtained when stacking changes from ABC to ABA by first separating the eight compounds into two groups according to their structural models of functionalization (Table \ref{tab1}). Sc$_{2}$NF$_{2}$, Sc$_{2}$NO$_{2}$, Cr$_{2}$NF$_{2}$ and Mn$_{2}$NF$_{2}$ are the four MXenes that stabilize in the same structural model HH when the stacking pattern changes. Inspecting Table \ref{tab:Table2}, we find that in all four cases, the magnetic ground state changes upon changing the stacking from ABC to ABA. The NM ground states in ABC stacking change to FM ones for both Sc$_{2}$NT$_{2}$ MXenes. In case of the other two, AFM3 ground states in ABC change to FM( for Mn$_{2}$NF$_{2}$) and AFM1(for Cr$_{2}$NF$_{2}$). In these two compounds, the magnetic moments of the transition metal constituents undergo insignificant changes. Significant changes in the electronic ground states are observed as stacking changes. In the ABA stacking, Sc$_{2}$NF$_{2}$ is a spin gapless semiconductor (Figure \ref{Fig:3}(a)) with total magnetic moment $ 1 \mu_{B}$ while Mn$_{2}$NF$_{2}$ is a half-metal (Figure \ref{Fig:3}(o))with a total moment of $9 \mu_{B}$. On the other hand, upon stacking change, Sc$_{2}$NO$_{2}$ becomes a near half-metal (Figure \ref{Fig:3}(c))with near integer total moment of $0.9 \mu_{B}$. The electronic ground state of Cr$_{2}$NF$_{2}$ remains semiconducting upon change in stacking. More robustness of magnetic and electronic ground states are observed in some of the four remaining MXenes, Ti$_{2}$NF$_{2}$,V$_{2}$NF$_{2}$, V$_{2}$NO$_{2}$ and Cr$_{2}$NO$_{2}$ that stabilize in CC configurations when the stacking pattern is ABA. V$_{2}$NF$_{2}$ and Cr$_{2}$NO$_{2}$ have the same magnetic ground states as those found in ABC stacking; the total and transition metal atomic magnetic moments in the later case remain completely unchanged. The electronic ground state of Cr$_{2}$NO$_{2}$ remains a half-metal too (Figure \ref{Fig:3}(k)). The electronic ground states of V$_{2}$NT$_{2}$ and Ti$_{2}$NF$_{2}$ change from semiconductor to metallic and near half-metal, respectively (Figure \ref{Fig:3}(e),(g),(i)). The atomic magnetic moments in these compounds change substantially when the stacking changes from ABC to ABA.   
\subsubsection{Understanding changes in magnetic moments and electronic ground states, with stacking}
The magnetic and electronic properties of the MXenes can be analyzed from the splitting of $d$-orbitals due to the crystal fields of the anions, their subsequent arrangements, and filling based upon associated local symmetry. In case of compounds subscribing to the HH structural model, the local $D_{3d}$ crystal field splits $d$-levels of M atoms of M$_{2}$NT$_{2}$ into doubly degenerate $e_{1} (d_{xz},d_{yz})$, single $a_{1} ( d_{z^{2}})$ and doubly degenerate $e_{2}(d_{x^{2}-y^{2}},d_{xy})$ in ascending order of energy. The stacking is not supposed to matter as long as the compounds adhere to the same structural model leading to the same local symmetry around M. We find this is indeed true for Cr$_{2}$NF$_{2}$ and Mn$_{2}$NF$_{2}$. The unequal magnetic moment on two inequivalent Cr atoms can be explained in the following way: each Cr donates 1 electron to N and 1 to the F attached to its surface. Since N requires another electron, one of the Cr has to donate from its share. In this case, it is Cr$^{II}$. As a result, ideally Cr$^{I}$(Cr$^{II}$) should have a magnetic moment of 4(3)$\mu_{B}$ per atom. From Table \ref{tab:Table2}, we find that the calculated values (in either stacking) are close to these. For a given spin band, the arrangement of electrons among $d$ bands of Cr$^{I}$ (Cr$^{II}$)should be $e_{1}^{2}a_{1}^{1}e_{2}^{1}(e_{1}^{2}a_{1}^{1}e_{2}^{0})$. That the two Cr atoms are inequivalent with regard to their oxidation states can be understood from their atom projected densities of states (Figure S7, supplementary information) for both stacking types. By comparing the densities of states of Cr$^{I}$ (Figure S7 (b), supplementary information) and Cr$^{II}$ (Figure S7(d), supplementary information), we find that while one spin band in Cr$^{I}$ is completely empty, the other is nearly full. It is not so for Cr$^{II}$. The same features are observed in ABA stacking (Figure S7(a),(c), supplementary information). In Figures S17 and S18, supplementary information, we show the densities of states associated with different $d$-orbitals of Cr$^{I}$ and Cr$^{II}$ respectively. As mentioned above, we find the $a_{1}$ orbitals for one-spin is completely full while other two orbitals are nearly full for Cr$^{I}$. Significant electronic re-distribution happens between $e_{1}$ and $e_{2}$ due to hybridizations with N and F states. In the case of Cr$^{II}$, $e$ states lie deeper in energy, signifying the deficit of an electron as compared to Cr$^{I}$. In case of Mn$_{2}$NF$_{2}$, two Mn atoms have identical magnetic moment of 4.5 $\mu_{B}$, irrespective of the stacking pattern. In this compound, one of the Mn should have 5 electrons while the other should have 4 electrons after donating to the F and N atoms. This should lead to unequal moments of 5 $\mu_{B}$ and 4 $\mu_{B}$. However, upon inspecting the atom-projected densities of states (Figure S9, supplementary information) and Mn $d$-orbital-projected densities of states (Figure S21, supplementary information), we find that contributions of both Mn are identical, with one of the spin bands near full. Like Cr$_{2}$NF$_{2}$, the $a_{1}$ states are full while the $e$ orbitals are near full. Compared to the ABC stacking, in ABA stacking, the states are pushed towards lower energies. The departure of Mn moments from integer values is due to the re-distribution of electrons among the $d$-orbitals as a consequence of strong hybridizations with anions. The delocalized states of F and N, particularly in ABC stacking, suggest significant hybridization. 

The picture is very different in case of Sc$_{2}$NF$_{2}$ and Sc$_{2}$NO$_{2}$. Figures S2, S3, S10, and S11, supplementary information show that the exchange splitting in these compounds in ABC stacking is weak, leading to an NM ground state. Sc has a valence electron configuration of $4s^{2}3d^{1}$. In Sc$_{2}$NF$_{2}$, after donating 5 electrons to N and two F atoms, there is one extra electron in the system. In ABA stacking, this is shared by the two Sc atoms, leading to an equal amount of magnetic moment on them. Change in stacking from ABC to ABA enhances the exchange splitting. The Sc-N bond lengths increase by 2\% , localizing the unpaired electron. This splits the majority band with a gap right at the Fermi level. The minority band states near the Fermi level in ABC now move to the unoccupied part of the energy spectra. The orbital-projected densities of states show that the peak in the majority band around -1 eV in ABA has contributions from all three $d$-orbitals(Figure S10 (a),(c),(e)). Sc$_{2}$NO$_{2}$, on the other hand does not have any unpaired $d$ electron left in the system. It rather has an unpaired N electron. In ABC stacking, the Sc bands of both spins are thus fully occupied, leading to an NM state. In ABA stacking, Sc atoms do not contribute to the magnetic moment. The unpaired N leads to a spin polarisation in N with a net magnetic moment. Subsequently, the ground state magnetic structure in these two compounds in ABA stacking is FM. 

The remaining four MXenes stabilize in the CC structural model in ABA stacking. Except Cr$_{2}$NO$_{2}$, the atomic magnetic moments in the other three compounds, Ti$_{2}$NF$_{2}$, V$_{2}$NF$_{2}$ and V$_{2}$NO$_{2}$, change considerably with change in stacking. In the CC structural model, the local crystal field symmetry surrounding the transition metal atoms is trigonal C$_{3v}$. In this case, the $d$ orbitals split into $a_{1},e_{1}$ and $e_{2}$ in the ascending order of energy. The changes in the atomic magnetic moments can be understood accordingly. To do this, we first consider Cr$_{2}$NO$_{2}$ where the two Cr atoms are equivalent. Each Cr donates 2 electrons to the corresponding O atom and 1 to the N atom out of its six valence electrons. Thus, each Cr atom is left with three unpaired electrons. However, N requires one more electron, which is contributed by both Cr, resulting in a magnetic moment of 5 $\mu_{B}$ for the system. The electron sharing between the cations and anions is strong, resulting in polarisations in N and O atoms in either stacking. This is evident from the atom-projected densities of states (Figure S8, supplementary information). In ABC stacking, the exchange splitting in the anions is nearly equal, while in ABA stacking, exchange splitting in N is greater than O, as is seen in Figure S8, (i). In the ABA stacking, the N majority states deplete in the energy range -1.0 to -3.0 eV when compared with O densities of states. The arrangement of electrons in various $d$ orbitals can be understood from the orbital-projected densities of states (Figure S20, supplementary information). In ABC stacking, we can see that the minority spin band of Cr is nearly empty. The electrons are distributed among the three orbitals, and none of them are fully occupied. There is significant hybridization between the majority states of Cr $a_{1}$  and $p$ orbitals of the O and N. In the ABA stacking, the lowest-lying $a_{1}$ majority band is full, and the minority band is empty. This is not so with the higher-lying $e$ orbitals. As such, there is little difference in the overall filling of electrons in energy levels when the stacking changes. 

Large differences in the atomic magnetic moments of the transition metal atoms as the stacking changes are observed for Ti$_{2}$NF$_{2}$ and V$_{2}$NO$_{2}$. In the case of Ti$_{2}$NF$_{2}$, each Ti donates 1 electron to F and 1 to N from its four valence electrons. From the remaining total of 4 electrons among them, 1 electron is to be donated to N. Ideally, one of the Ti should have a magnetic moment of 2$\mu_{B}$ while the other should have a moment of 1$\mu_{B}$. In the ABC stacking, we find the inequivalent Ti atoms; Ti$^{I}$ is having a moment of 1.3 $\mu_{B}$ and Ti$^{II}$ of 1$\mu_{B}$. Once again, the atom projected densities of states (Figure S4 (b),(d),(f),(h),(j)) suggests that only one of the spin bands of each Ti atom is partially full while the other one is either completely (in case of Ti$^{I}$) or near completely (in case of Ti$^{II}$) empty. The variations in the exchange splitting due to this are responsible for unequal magnetic moments of Ti. Upon inspecting the orbital-projected densities of states (Figure S12-S13, (b),(d),(f), supplementary information), we find that the major difference among Ti$^{I}$ and Ti$^{II}$ is that in the latter case the $a_{1}$ orbitals of both spin bands are almost empty. The $e$ states for Ti$^{I}$ are spread out up to nearly -2 eV, while those of Ti$^{II}$ are more localized near -2 eV. There is significant hybridization leading to electron sharing among Ti$^{I}$ and the anions, as the contributions from all of them happening in the same energy window. Different sequences of $d$-orbital levels and subsequent filling are responsible for the substantial reduction of Ti moments in ABA stacking (Figure S4 (a),(c),(e),(f),(g) and S12-S13 (a),(c),(e), supplementary information). Here, the lowest lying $a$ orbitals of both spins are partially filled, the $e_{1}$ is completely empty, and $e_{2}$ has partial filing in both spin bands. The smaller exchange splittings, thus, reduce the Ti moments in ABA stacking. V$_{2}$NO$_{2}$ has identical situation. Atomic magnetic moments and their changes with stacking change, too, can be understood in a similar way (Figures S6, S16-S17, supplementary information). In the case of V$_{2}$NF$_{2}$, though the changes in magnetic moments with stacking are not as large as observed in these two compounds, they are substantial. The five unpaired V electrons are distributed asymmetrically among two V atoms, making them non-equivalent with respect to magnetic moments in ABC stacking. Strong cation-anion hybridization and localized states are the prominent features of the electronic structure of this compound in ABC stacking (Figure S5, S14-S15 (b),(d),(f),(h),(j)). The split $d$-levels have one spin band empty, and each one of the other spin is partially filled. In ABA stacking, the states are less localized. The minority spin bands associated with the split $d$ levels are not completely empty. This explains the reduction in moments with respect to the moments obtained in ABC stacking. The electrons totally occupy the $a_{1}$ and $e_{2}$ levels only as $e_{1}$ is completely empty. 

The changes in the electronic ground states with changes in the stacking are intricately related to the above-discussed patterns of orbital occupation and anion-cation hybridizations. For example, V$_{2}$NF$_{2}$ has an identical magnetic ground state but is a semiconductor(metal) in the ABC(ABA) stacking. The localized anion-cation orbital overlap is responsible for a semiconducting ground state. Compared to this, there is much more itinerancy in the $d$ states when the stacking is ABA. The anion $d$-cation $p$ interaction is weak, too (Figure S5, supplementary information), leading to a metallic ground state. The changes in the electronic ground states with stacking for other compounds can be understood this way. 

A noteworthy point here is that except in Cr$_{2}$NF$_{2}$, ABA stacking renders all transition metal constituents equivalent. It is not so in the case of ABC stacking. Apart from Cr$_{2}$NF$_{2}$, Ti$_{2}$NF$_{2}$, V$_{2}$NF$_{2}$ and V$_{2}$NO$_{2}$ have two inequivalent M as can be made out from unequal magnetic moments on them. This is due to the inhomogeneous environment around the M atoms in these systems. Although the chemical species around them are identical, the variations in the anion-cation bond distances introduce inhomogeneity in the environment, leading to different exchange splittings of inequivalent atoms. For example, in V$_{2}$NO$_{2}$, V-O(V-N) bond distances are 2.0 \AA(1.92 \AA) and 2.19 \AA(2.13 \AA) in ABC stacking. A difference of 9.5 \% and 11 \% in the V-O and V-N bond distances associated with two V, respectively, can be correlated with large differences in their magnetic moments. V$^{II}$-N/O bonds are much larger than the V$^{I}$-N/O bonds. As a result, V$^{II}$ states are more localized with less hybridization with anions and a smaller exchange splitting (Figure S6, supplementary information). In ABA stacking, there is uniformity in both bond distances; the V-O(V-N) bond distance is 1.96 \AA(2.11 \AA). In Cr$_{2}$NF$_{2}$, Cr-F (Cr-N) bonds vary by 1.4\%(6.4\%) in ABC stacking. The numbers are 1.9\% (Cr-F) and 4.7\% (Cr-N) in ABA stacking. Larger variations in Cr-N bond lengths affect cation-anion hybridizations and, subsequently, the magnetic moments on inequivalent atoms.  

\begin{table*}
\caption{\label{tab:Table2}The magnetic properties of the 8 MXenes in both stacking pattern. M$^{I}$ and M$^{II}$ are the magnetic moments (in $\mu_{B}$/atom)of two non equivalent transition metal atoms.M$^{N}$ is the magnetic moment (in $\mu_{B}$/atom) of Nitrogen atom . $T_{C}$/$T_{N}$ is the magnetic transition temperature in K. MAE is the magnetic anisotropy energy per unit cell in $ \mu eV $ . }
\begin{adjustbox}{width=1\textwidth}
\small
\begin{tabular}{c|cclllcr}
 \hline
 compound & stacking type & Magnetic Ground state &   M$^{I}$  & M$^{II}$ & M$^{N}$ & $T_{C}$/$T_{N}$ & MAE\\
 \hline
  & ABA  & FM & 0.44 & 0.44 & 0.05 &16 & -2.28 \\
 Sc\textsubscript{2}NF\textsubscript{2} & & & & & & \\
 \cline{2-8}\\
 & ABC & NM & 0 & 0& 0 & - & - \\
 \hline
 & ABA & FM & 0 & 0 & 0.56 & 30 & 23.80 \\
 Sc\textsubscript{2}NO\textsubscript{2} & & & & & &\\
  \cline{2-8}\\
 & ABC & NM(FM$^{\cite{sc2no2}}$) & 0 & 0 & 0.03(1.0$^{\cite{sc2no2}}$)  & - & - \\
 \hline
 & ABA & FM & 0.40 & 0.40 & 0.04 & 35 & 19.38 \\
 Ti\textsubscript{2}NF\textsubscript{2} & & & & & &\\
  \cline{2-8}\\
 & ABC & AFM2(AFM2$^{\cite{kumar2017tunable}}$) & 1.3(1.3$^{\cite{kumar2017tunable}}$) & 1.0(1.0$^{\cite{kumar2017tunable}}$)& 0.02 & 50 & -4.76 \\
 \hline
 & ABA & AFM2 & 1.78 & 1.78 & 0.07 & 350 & -148.40\\
 V\textsubscript{2}NF\textsubscript{2}& & & & & & \\
  \cline{2-8}\\
 & ABC & AFM2(AFM2$^{\cite{kumar2017tunable}}$, AFM1$^{\cite{mxene-nanomaterials}}$) & 2.4(2.5$^{\cite{kumar2017tunable}}$, 2.7$^{\cite{mxene-nanomaterials}}$) & 2.0(2.0$^{\cite{kumar2017tunable}}$, 2.1$^{\cite{mxene-nanomaterials}}$) & 0.01 & 90  & 1.5 \\
 \hline
 & ABA &  AFM2 & 0.50 & 0.50 & 0.03 &  45 & 6.4 \\
 V\textsubscript{2}NO\textsubscript{2} & & & & & &\\
  \cline{2-8}\\
 &ABC & AFM1(AFM2$^{\cite{kumar2017tunable}}$, AFM1$^{\cite{mxene-nanomaterials}}$) & 1.75(1.8$^{\cite{kumar2017tunable}}$, 1.9$^{\cite{mxene-nanomaterials}}$) & 1.11(1.0$^{\cite{kumar2017tunable}}$, 1.2$^{\cite{mxene-nanomaterials}}$) &  0.00 & 60 & -55.5 \\
  \hline\\
 & ABA & AFM1 & 3.70 & 3.10 & 0.07 & 50 & 148.50 \\
 Cr\textsubscript{2}NF\textsubscript{2} & & & & & & \\
  \cline{2-8}\\
 &ABC & AFM3(AFM2$^{\cite{kumar2017tunable}}$, AFM1$^{\cite{mxene-nanomaterials}}$) & 3.65(3.7$^{\cite{kumar2017tunable}}$, 3.9$^{\cite{mxene-nanomaterials}}$) & 3.11(3.0$^{\cite{kumar2017tunable}}$, 3.1$^{\cite{mxene-nanomaterials}}$) & 0.0 & 60 & -48.51 \\
 \hline
 & ABA & FM & 2.91 & 2.91 & 0.45 & 90 & -166.40 \\
 Cr\textsubscript{2}NO\textsubscript{2} & & & & & &\\
  \cline{2-8}\\
 & ABC & FM(FM$^{\cite{kumar2017tunable}}$) &
  2.87(2.8$^{\cite{kumar2017tunable}}$)  & 2.87(2.8$^{\cite{kumar2017tunable}}$) & 0.23 & 200 & 36.95\\
  \hline
 & ABA & FM & 4.50 & 4.50 & 0.37 & 50 & -13.50\\
 Mn\textsubscript{2}NF\textsubscript{2} & & & & & &\\
  \cline{2-8}\\
 & ABC & AFM3(AFM1$^{\cite{kumar2017tunable}}$) & 4.5(4.5$^{\cite{kumar2017tunable}}$) & 4.5(4.5$^{\cite{kumar2017tunable}}$) & 0.05 & 60 & 19.87 \\
 \hline
\end{tabular}
\end{adjustbox}
\end{table*}

\subsubsection{Understanding the origin of magnetic ground states and their stacking dependence}
\begin{figure}
    \includegraphics[height=7cm, width=8.00 cm]{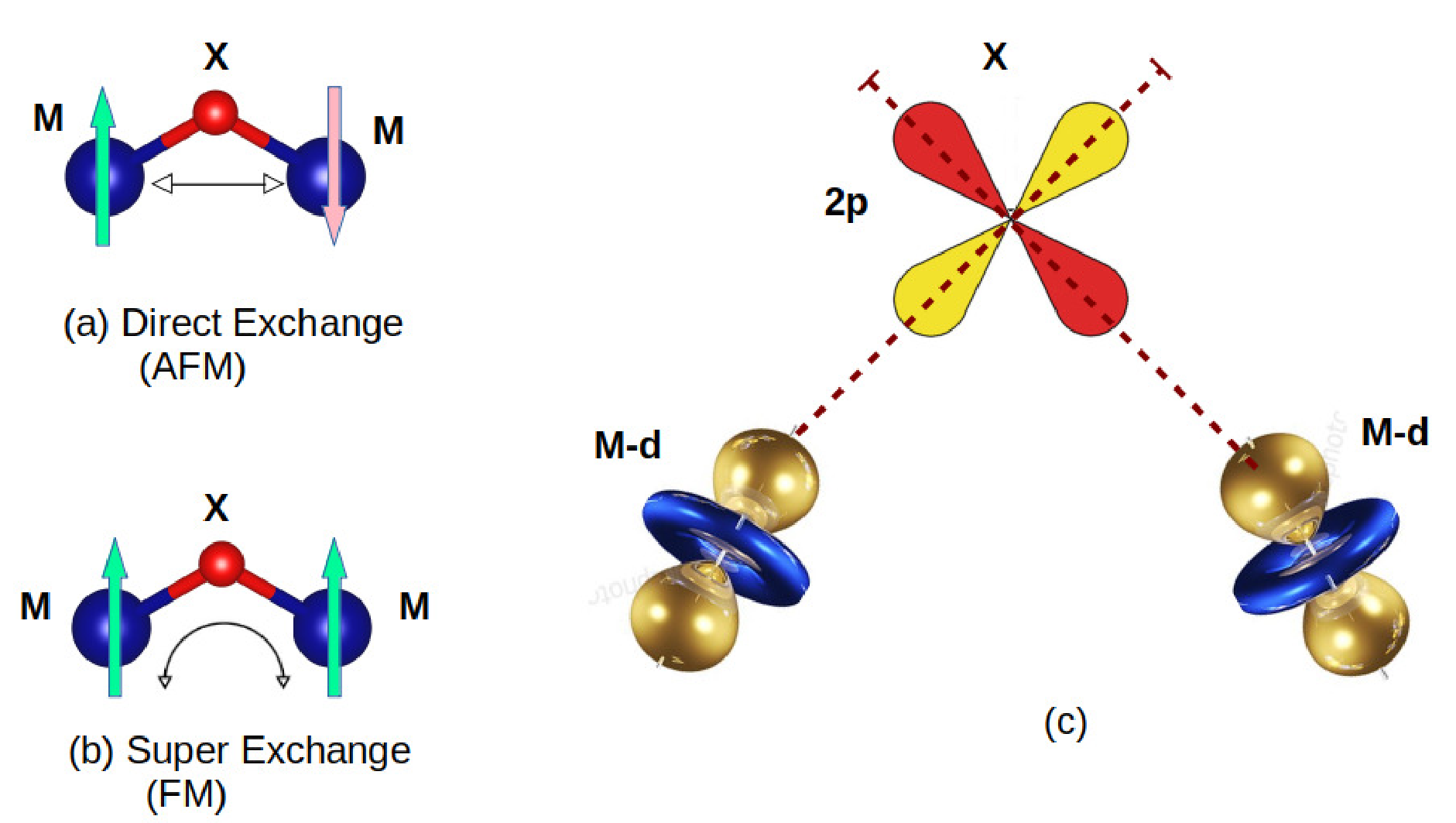}
     \caption{Various exchange mechanisms in the MXenes considered are shown. (a) and (b) show the direct exchange and superexchange, respectively. (c) shows the schematics of cation-anion-cation orbital overlaps in superexchange where same cation $d$-orbitals interact with different $p$ orbitals of the same anion.}
    \label{Fig:4}
\end{figure}

For all MXenes considered in this work, the M-X-M bond angle is close to 90\degree. This allows the $d$-orbitals of transition metal atoms to overlap with different $p$-orbitals of the anions. According to Goodenough Kanamori rule \cite{goodenough}, this superexchange results in an FM arrangement of the M atoms (Figure \ref{Fig:4}(a),(c)). On the other hand, direct exchange between M atoms (Figure \ref{Fig:4}(b)) leads to an AFM arrangement. The competition between these two exchange mechanisms often determines the magnetic ground state configuration. It depends on the degree of occupancy of the $d$ orbitals and/or the inter-atomic distances. For the MXenes considered here, the origin of a particular magnetic ground state in a given stacking can, by and large, be explained this way. 

In Ti$_{2}$NF$_{2}$, the origin of an FM ground state in ABA stacking is the nearly filled $a_{1}$ states that weaken the AFM exchange interaction, making the superexchange dominating. In ABC stacking, half-filled $e_{1}$ states promote direct exchange as hopping through spin flipping is allowed. This results in the AFM2 ground state. Since  $e_{1}$ orbitals of opposite spin bands coming from inequivalent Ti atoms lying on the same surface are half-filled, the dominating direct exchange is between the inequivalent atoms leading to the AFM2 configuration. The origin of an AFM ground state in V$_{2}$NF$_{2}$ with ABC stacking can be explained similarly. Moreover, the inter-atomic distance between inequivalent V atoms in the ABC stacking is 2.92 \AA. With ABA stacking, the inter-atomic distance between V atoms reduces further to 2.8 \AA. Typically, inter-atomic distances beyond 3.0 \AA make the direct exchange weaker. Thus, in this compound, direct exchange is promoted over superexchange. The inter-atomic distance is crucial in deciding the ground state magnetic structure of V$_{2}$NO$_{2}$ as well. In ABA stacked V$_{2}$NO$_{2}$, the distance between nearest neighbor V atoms lying in the same plane is only 2.78 \AA. This results in a dominant direct exchange and an AFM2 ground state. With the change of stacking to ABC, the distance between inequivalent atoms lying in the same plane increases to 2.97 \AA. This weakens the direct exchange. The dominant superexchange results in an in-plane FM arrangement. However, the inter-planar distance being smaller than this (2.93 \AA), the direct exchange prevails along $c$-direction leading to an AFM1 ground state magnetic configuration. In Cr$_{2}$NF$_{2}$, ABA(ABC) stacking leads to an AFM1(AFM3) ground state. Here, the origin of this difference can be traced back to the intra-planar and inter-planar distances between Cr atoms. The intra-planar Cr-Cr distance is greater in ABA stacking (3.18 \AA in ABA and 3.08 \AA in ABC). Consequently, the direct exchange (superexchange) in ABA is less (more) than that in ABC. This leads to an in-plane FM(AFM) configuration in ABA(ABC). On the other hand, the inter-planar Cr-Cr distance in ABA is much smaller than that in ABC (2.49 \AA  in ABA and 3.09  \AA  in ABC). As a result, in ABA(ABC), the dominant inter-planar exchange is a direct exchange (superexchange) giving rise to AFM1 (AFM3) ground state. In Mn$_{2}$NF$_{2}$, there is an AFM to FM transition of the ground state magnetic configuration as stacking changes from ABC to ABA. Here, the ground state is determined by the competition between intra- and inter-planar interactions. In ABA, the inter-planar Mn atoms are the first and third nearest neighbors. The nearest neighbor distance being larger than 3.0 \AA, superexchange is promoted. The second, fourth, and sixth neighbors lie in the same plane. However, the inter-planar distances being close to 3.0 \AA and higher, superexchange dominates. Strong anion-cation overlap is also observed In  ABA (Figure S9, supplementary information). In the ABC stacking, the overlap of Mn $d$ and F $p$ orbitals is relatively weak, as can be seen from the atom-projected densities of states in the region of -2 to -4 eV (Figure S9, supplementary information). Naturally, it leads to an AFM ground state. The occurrence of FM ground state in Cr$_{2}$NO$_{2}$, irrespective of stacking pattern, can be similarly explained from the atom projected densities of states (Figure S8, supplementary information) which show strong anion $p$-cation $d$ overlap suggesting dominant superexchange mechanism.   

\begin{figure*}
    \includegraphics[height=7 cm, width=18.00 cm]{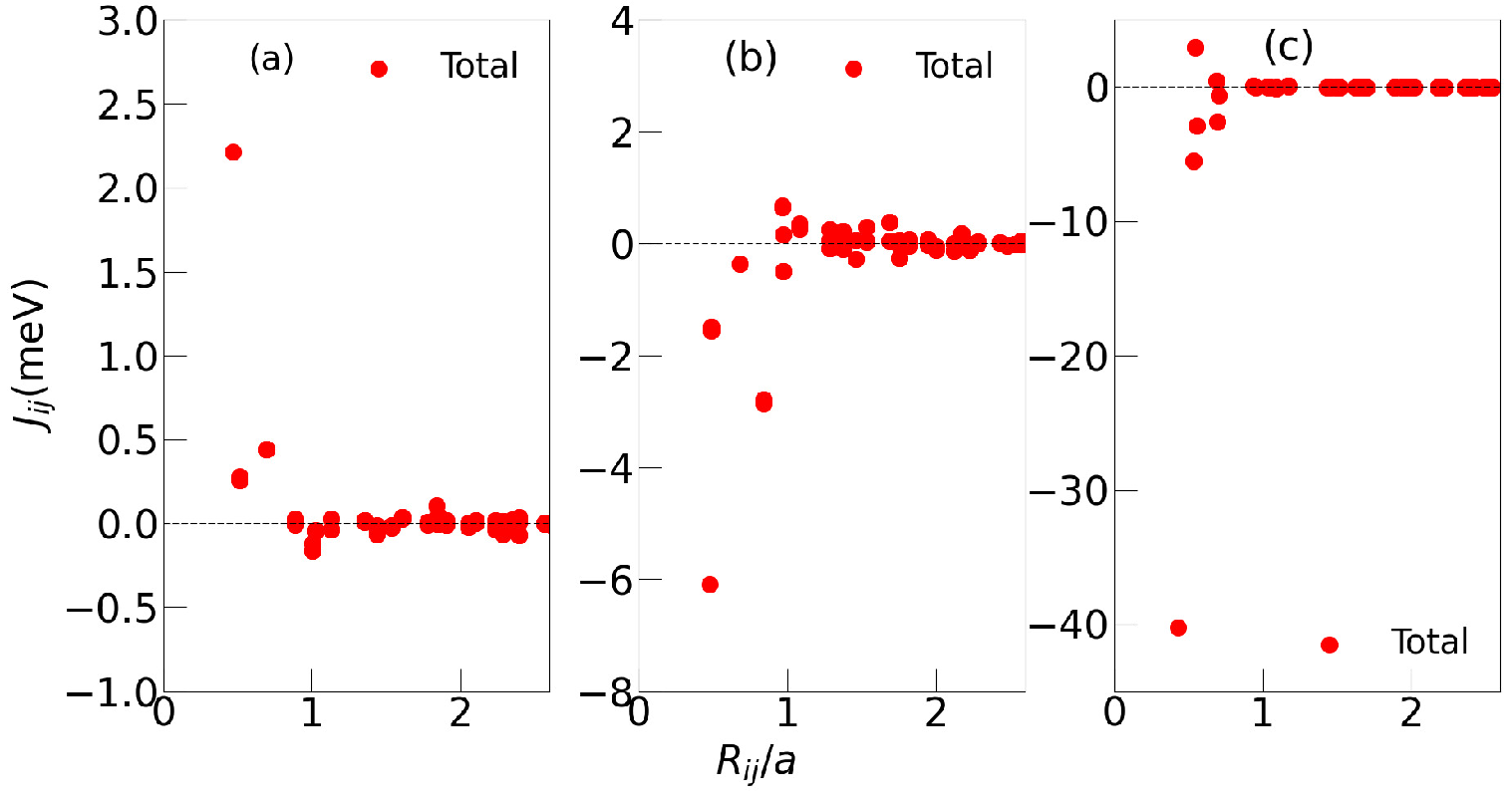}
 \caption{Inter-atomic exchange parameters as a function of inter-atomic distance for (a) Ti\textsubscript{2}NF\textsubscript{2} and (b) V\textsubscript{2}NO\textsubscript{2} and (c) Cr\textsubscript{2}NF\textsubscript{2} in ABA stacking.}
    \label{Fig:5} 
\end{figure*}
\subsection{Interatomic exchange parameters, magnetic transition temperatures and their dependencies on stacking pattern}
\begin{figure*}
    \includegraphics[height=7 cm, width=18.00 cm]{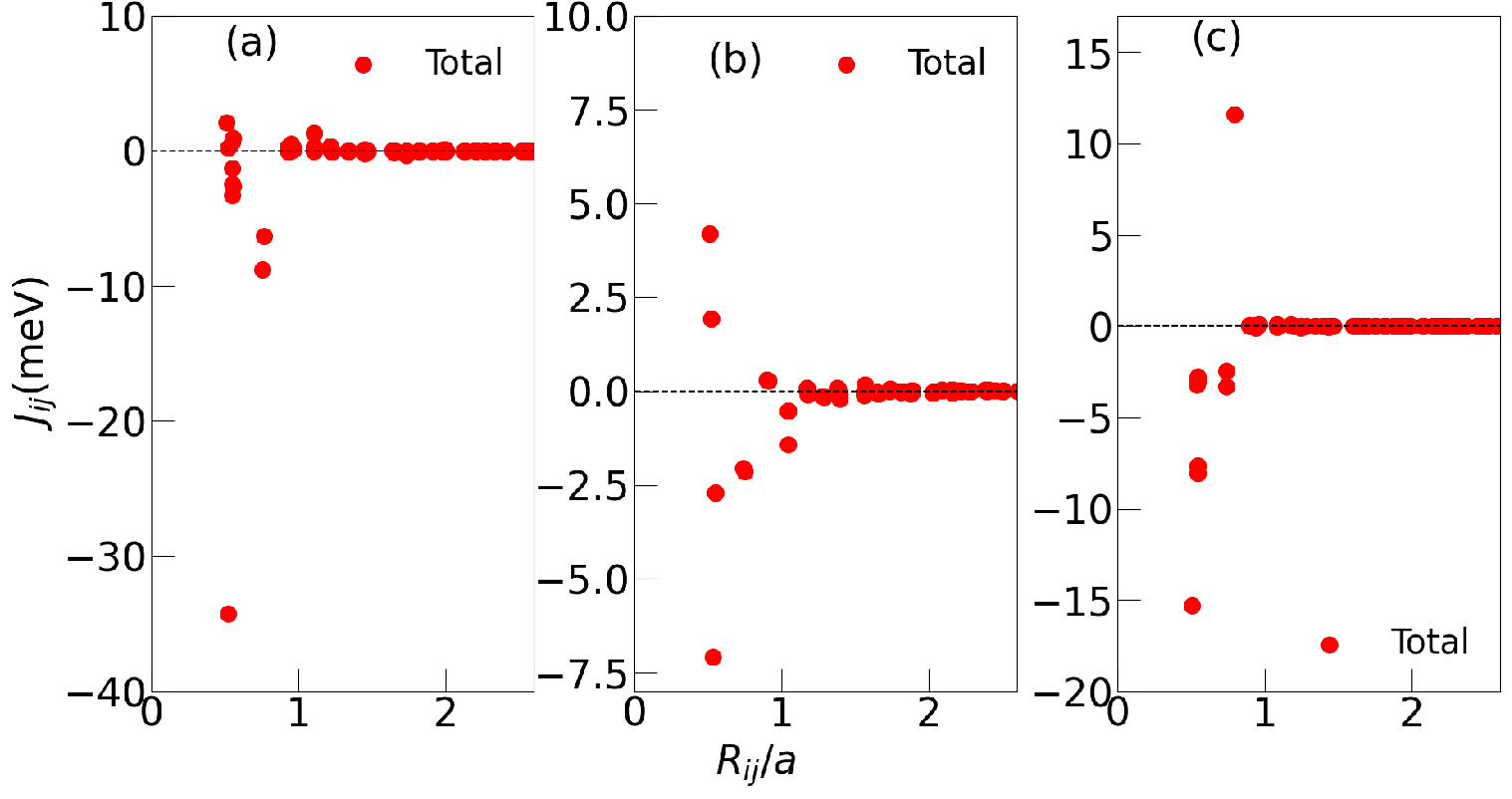}
 \caption{Inter-atomic exchange parameters as a function of inter-atomic distance for (a) Ti\textsubscript{2}NF\textsubscript{2} and (b) V\textsubscript{2}NO\textsubscript{2} and (c) Cr\textsubscript{2}NF\textsubscript{2} in ABC stacking.}
    \label{Fig:6} 
\end{figure*}
 Figure \ref{Fig:5}(Figure \ref{Fig:6})  and S22(S23), supplementary information show the interatomic exchange interactions of the systems in ABA(ABC) type stacking. The calculated exchange interaction parameters corroborate the explanations for the obtained magnetic ground states given in the previous subsection. For ABA stacked Ti$_{2}$NF$_{2}$, the dominant interactions from the first three neighbors are ferromagnetic, with the nearest neighbor interaction having a strength of 2.21 meV. The interactions beyond the first three neighbors are negligibly small. This makes the magnetic ground state of this compound FM. The first few interactions in V$_{2}$NO$_{2}$ are AFM, with the strongest one of magnitude 6.05 meV coming from the nearest neighbor pairs. Here, the nearest neighbors lie in the same plane across a distance of 2.78 \AA. As discussed earlier, such a distance makes the dominant in-plane interaction AFM and favors an AFM2 ground state. In ABA stacked Cr$_{2}$NF$_{2}$ (Figure \ref{Fig:5}(c)), we find the nearest neighbor interaction strong AFM with a magnitude of -40 meV. Here, the nearest neighbor pairs are situated out-of-plane. The strong out-of-plane AFM interaction drives the ground state AFM1. 

In ABC stacking, the inequivalent Ti atoms of Ti$_{2}$NF$_{2}$ lie in the same plane. Strong AFM interaction from first neighbor pairs ($\sim$ 35 meV) (Figure \ref{Fig:6}(a))followed by AFM interactions from a few higher neighbors is the reason behind the AFM ground state. Since the nearest neighbors are the inequivalent atoms, the resulting AFM configuration is AFM2. In ABC stacked V$_{2}$NO$_{2}$, we find competing FM and AFM interactions in the first few neighbors. The nearest neighbor distance in the ABC stacked compound is now 2.93 \AA, connecting out-of-plane V atoms. This is substantially larger(2.78 \AA) than that in the ABA stacked compound. The in-plane distance of V atoms is 2.97 \AA. From Figure \ref{Fig:6}(b), we find the competing FM and AFM interactions. The out-of-plane nearest neighbors connect via strong AFM exchange, while the third and fourth neighbor interactions are FM with competitive magnitudes. The in-plane FM ordering and out-of-plane AFM ordering happen as an outcome. In Cr$_{2}$NF$_{2}$, ABC stacking puts the in-plane inequivalent atoms at the nearest neighboring distance. The strong AFM interactions among the nearest neighbor and a few subsequent higher-order pairs keep the in-plane interaction AFM, resulting in an AFM3 ground state. The changes in the exchange interactions and subsequent stabilization of the ground state magnetic configuration with changes in stacking for the other five MXenes can be interpreted similarly (Figures S22-23, supplementary information).

 The exchange interactions reported above are used to compute the magnetic transition temperatures. The results are given in Table \ref{tab:Table2}. Substantial changes in magnetic transition temperature upon a change in stacking pattern are observed in V$_{2}$NF$_{2}$ and Cr$_{2}$NO$_{2}$. For the former, ABA stacking led to a transition temperature of 350K, which is close to room temperature. For the latter, a transition temperature of 200K is obtained in ABC stacking. Such differences can be attributed to the changes in magnetic exchange parameters. Notably, most of the systems considered in this work have magnetic ordering temperatures substantially higher than CrI\textsubscript{3} \cite{huang2017layer}, the prototype 2D magnet. 
\subsection{Magnetic Anisotropy energy and its microscopic origin}
\begin{figure*}
    \includegraphics[height=14 cm, width=19.00 cm]{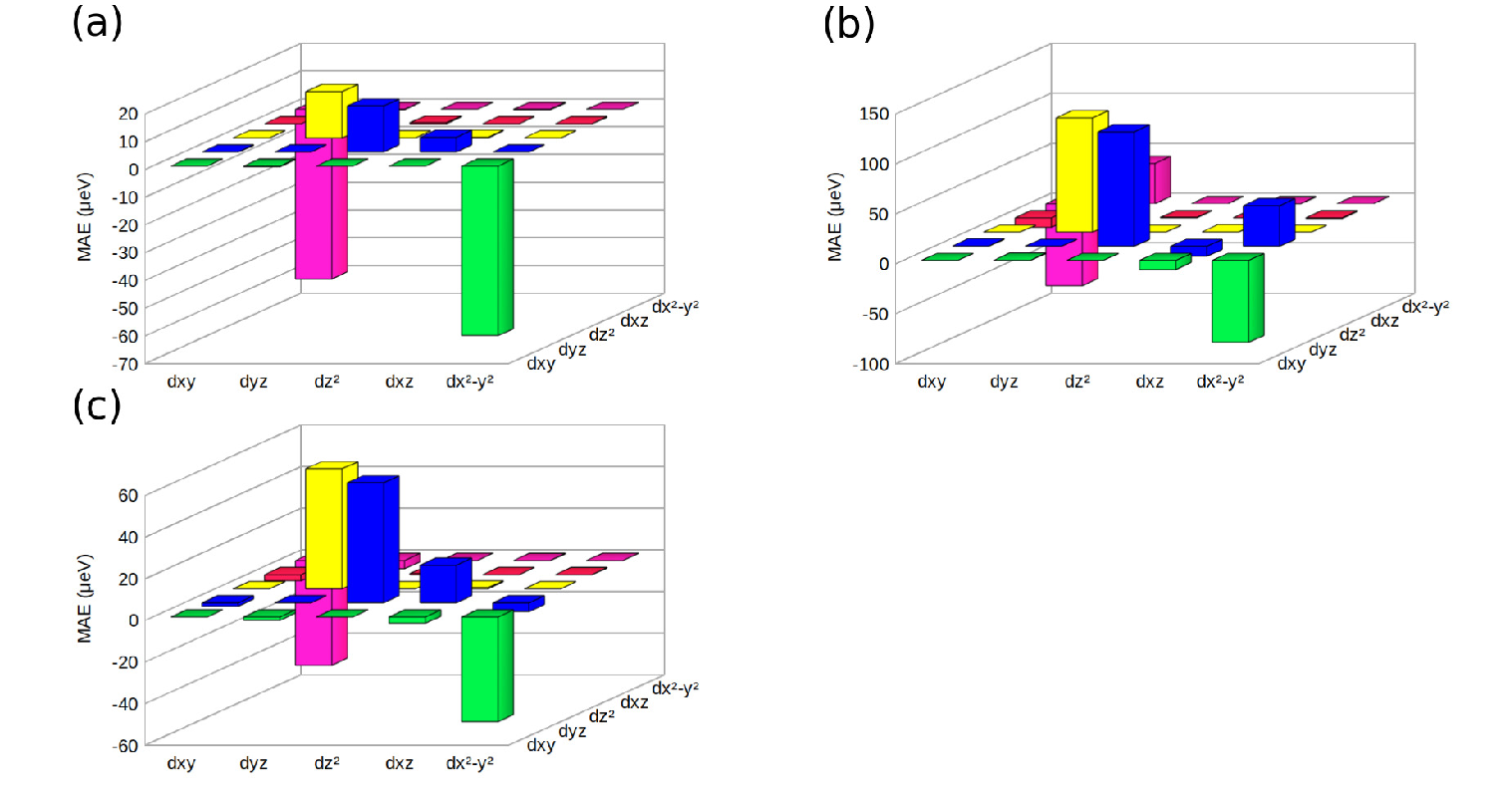}
 \caption{Contributions of various orbitals towards magnetic anisotropy energy of (a) Cr\textsubscript{2}NO\textsubscript{2},(b) Cr\textsuperscript{I} of Cr\textsubscript{2}NF\textsubscript{2} (c)Cr\textsuperscript{II} of Cr\textsubscript{2}NF\textsubscript{2} and (d) Sc\textsubscript{2}NO\textsubscript{2}. Results are for ABA stacking.}
    \label{Fig:7} 
\end{figure*}
 Magnetic anisotropy enables magnetism to survive in the 2D limit. It is also important from the perspective of device application as it stabilizes the magnetic long-range order in a given system. The magnetic anisotropy energy (MAE) can be  evaluated as:
 \begin{equation}
 E_{MAE} = E_{||}- E_{\perp}
 \end{equation} 
Here, $ E_{||} $ and  $E_{\perp}$ correspond to the total energy of the magnetic system calculated with the spins aligned parallel and perpendicular to the plane, respectively, taking spin-orbit coupling into account. The obtained values of MAE for our systems are given in Table \ref{tab:Table2}. A positive (negative) value of MAE indicates the existence of an easy axis perpendicular to the MXene surface(easy plane). 

The calculated values of MAE show that the nature of anisotropy, in-plane(MAE$<0$) or out-of-plane (MAE$>0$), in these MXenes, depends on the type of stacking. Quite a few MAE values are comparable to those obtained in well-known 2D magnets like CrBr$_{3}$, RhO$_{2}$, Cr$_{2}$Ge$_{2}$Te$_{6}$ and CrS$_{2}$ \cite{jiang2021recent,lee2022out,xiao2022novel,liu2022magnetic}. Significant differences of MAE when the stacking changes are observed in three cases: Cr$_{2}$NO$_{2}$ and V$_{2}$NF$_{2}$ where ABA stacking leads to very large in-plane anisotropy while ABC stacked compounds show out-of-pane anisotropy with MAE either very small(for V$_{2}$NF$_{2}$) or moderate (for Cr$_{2}$NO$_{2}$) and Cr$_{2}$NF$_{2}$ where ABA(ABC) stacked compound exhibits very large(moderate) out-of-plane(in-plane) anisotropy. In all the other MXenes where both stackings led to a magnetic order, the qualitative nature of anisotropy is observed to have changed with stacking.  

The changes in the qualitative and, in some cases, quantitative nature of the magnetic anisotropy can be understood from the calculation of MAE using second-order perturbation theory. Using perturbation theory, MAE can be estimated through spin-orbit (SOC) interactions between the occupied and unoccupied states of a system \cite{wang1993first,lee2022out}:
\begin{align}
\resizebox{\hsize}{!}{$E_{MAE} = \xi^{2}\sum_{o,u,\alpha,\beta} (-1)^{1-\delta_{\alpha \beta}}\frac{\mid \langle o^{\alpha}\mid L_{z}\mid u^{\beta}\rangle\mid^{2} - \mid \langle o^{\alpha}\mid L_{x}\mid u^{\beta}\rangle\mid^{2}}{\epsilon^{\beta}_{u} - \epsilon^{\alpha}_{o}}$}.
\end{align}
 $o^{\alpha}(u^{\beta})$ is the occupied(unoccupied) orbital for spin $\alpha$($\beta$), $\epsilon^{\alpha}_{o}$($\epsilon^{\beta}_{u}$) are the corresponding eigenenergies. $L_{z}$($L_{x}$) is the $z(x)$ component of the angular momentum operator. $\xi $ is the spin-orbit coupling constant. There are only five non-zero matrix elements connecting the $d$-orbitals: $ \langle d_{xz}\vert L_{z} \vert d_{yz} \rangle $, $ \langle d_{x^{2}-y^{2}}\vert L_{z} \vert d_{xy} \rangle $,$ \langle d_{xy}\vert L_{z} \vert d_{xz} \rangle $,$ \langle d_{z^{2}}\vert L_{x} \vert d_{yz} \rangle $, and $ \langle d_{x^{2}-y^{2}}\vert L_{x} \vert d_{yz} \rangle $.The maximum contributions to MAE are due to states closer to the Fermi level. Thus, the electronic structure of materials near the Fermi level becomes decisive. Also, the first(second) term in the numerator in Equation(2) leads to positive (negative) contributions towards MAE when the spins associated with the orbitals are parallel. For orbitals with anti-parallel spin orientations, the situation is exactly the opposite.
 \begin{figure*}
    \includegraphics[height=12 cm, width=19.00 cm]{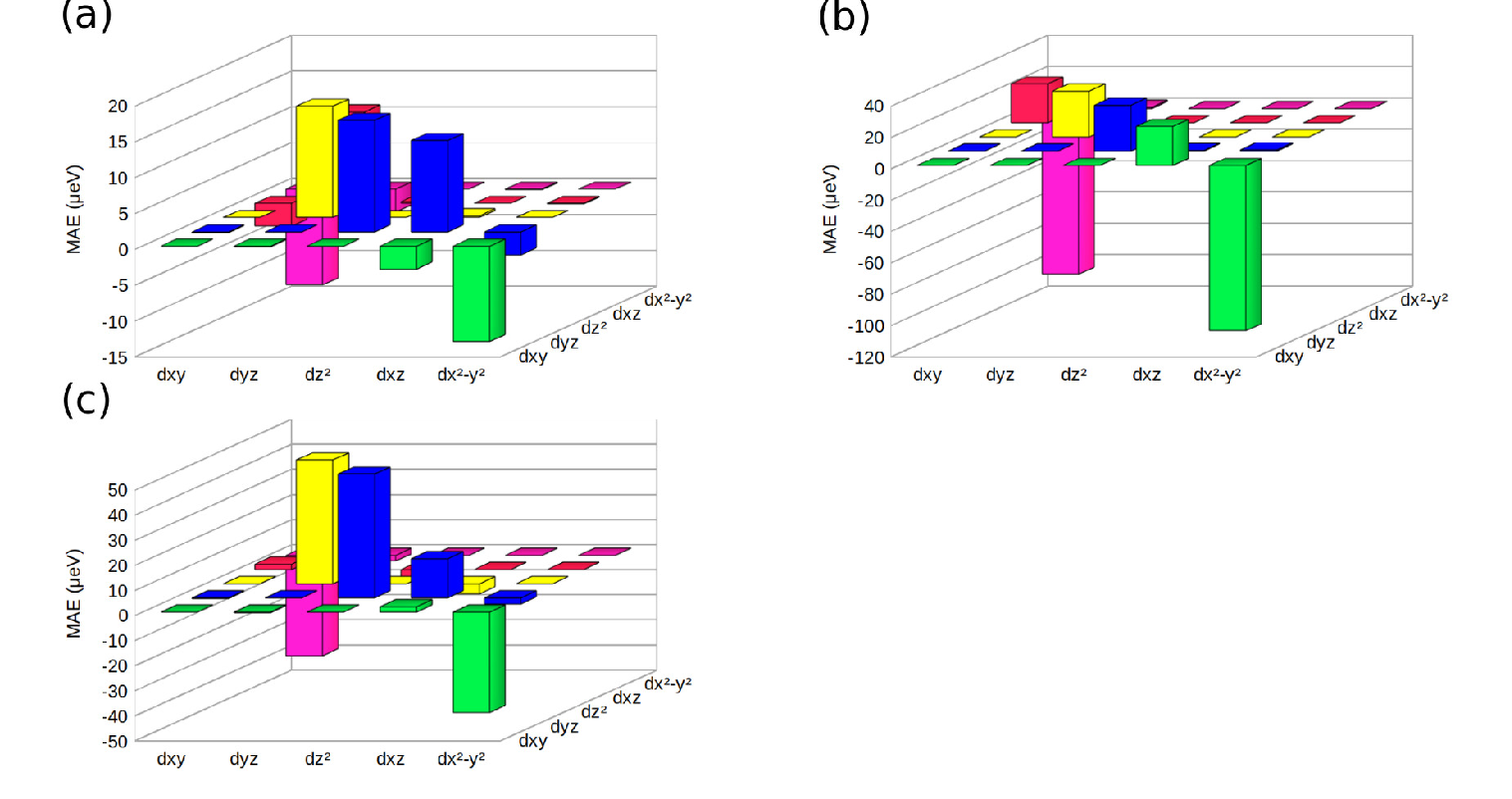}
 \caption{Contributions of various orbitals towards magnetic anisotropy energy of (a) Cr\textsubscript{2}NO\textsubscript{2},(b) Cr\textsuperscript{I} of Cr\textsubscript{2}NF\textsubscript{2} (c)Cr\textsuperscript{II} of Cr\textsubscript{2}NF\textsubscript{2}.The results are for ABC stacking. }
    \label{Fig:8} 
\end{figure*}

To understand the qualitative and quantitative nature of MAE and its changes with stacking, we consider Cr$_{2}$NO$_{2}$ and Cr$_{2}$NF$_{2}$ since these are the compounds where significant qualitative and quantitative changes occur when stacking changes. Contributions from the interactions of various pairs of $d$-orbitals in ABA (ABC) stacked compounds are presented in Figure \ref{Fig:7}(Figure \ref{Fig:8}).
Only the contributions from the $d$-orbitals of Cr atoms are considered, as the other contributions are negligible. The interactions of $d$-orbitals contributing to the MAE are represented as  $ d_{i}\otimes d_{j} $ interactions \cite{lee2022out}. We first consider the case of Cr\textsubscript{2}NO\textsubscript{2}. Results for Cr\textsubscript{2}NO\textsubscript{2} in ABA stacking (Figure \ref{Fig:7}(a)) suggest that the main contributions to MAE come from $ d_{xy}\otimes d_{x^{2}-y^{2}} $ and  $  d_{yz}\otimes d_{z^{2}} $ pairs. The contribution from  $ d_{xy}\otimes d_{x^{2}-y^{2}} $ is large and negative whereas the contribution from $  d_{yz}\otimes d_{z^{2}} $  is small and positive . Therefore, it is primarily due to  $ d_{xy}\otimes d_{x^{2}-y^{2}} $ contributions, Cr\textsubscript{2}NO\textsubscript{2} has a large and negative value of MAE in ABA stacking.

In the case of ABC stacked Cr\textsubscript{2}NO\textsubscript{2}, the main contributions to MAE come from three pairs of interactions represented as $ d_{xy}\otimes d_{x^{2}-y^{2}} $, $  d_{yz}\otimes d_{z^{2}} $, and $  d_{yz}\otimes d_{xz}$ (Figure \ref{Fig:8}(a)). $  d_{yz}\otimes d_{z^{2}} $ , and $  d_{yz}\otimes d_{xz} $  have positive contributions whereas the contribution from  $ d_{xy}\otimes d_{x^{2}-y^{2}} $ is negative . The positive contributions of $  d_{yz}\otimes d_{xz} $  is almost equal to the negative contribution from $ d_{xy}\otimes d_{x^{2}-y^{2}} $. Thus, it is the positive contribution from $  d_{yz}\otimes d_{z^{2}} $ that is responsible for a modest and positive MAE in Cr\textsubscript{2}NO\textsubscript{2}. 

Since $ d_{xy}\otimes d_{x^{2}-y^{2}} $ coupling is the major contributor, irrespective of stacking pattern, further analysis of this contribution and its changes with stacking is required. The $ d_{xy}\otimes d_{x^{2}-y^{2}} $ interaction has four components: (a) $ (o^\uparrow \vert u^\uparrow ) $ indicating SOC interaction between occupied and unoccupied spin-up orbitals, (b) $ (o^\uparrow \vert u^\downarrow ) $ implying SOC interaction between occupied spin-up and unoccupied spin-down orbitals (c) $ (o^\downarrow \vert u^\uparrow ) $ signifying SOC interaction between occupied spin-down and unoccupied spin-up states and (d) $ (o^\downarrow \vert u^\downarrow ) $  coming from SOC interaction between occupied and unoccupied spin-down states. Contributions from $ (o^\uparrow \vert u^\uparrow ) $ and   $ (o^\downarrow \vert u^\downarrow ) $ are $ > 0 $  whereas  that from $ (o^\uparrow \vert u^\downarrow ) $ and  $ (o^\downarrow \vert u^\uparrow ) $ are $ < 0 $. The origin of overwhelmingly larger contributions from $ (o^\uparrow \vert u^\downarrow ) $ in ABC lies in their electronic structures. From the $d$-orbital resolved densities of states of both ABA and ABC stacked Cr\textsubscript{2}NO\textsubscript{2}(Figure S20, supplementary information), we observe that in ABA stacking, states in the unoccupied spin-down channel of $ d_{xy}/d_{x^{2}-y^{2}}$  are nearer to Fermi level than in the ABC stacking. Consequently, the contributions of $o^\uparrow $ and $ u^\downarrow $ are less in ABC stacking. Also, contributions from $ o^\downarrow $ and $ u^\uparrow $ are negligible because the corresponding states are far from the Fermi level. This explains the reason behind large(moderate) and negative(positive) contribution from $ d_{xy}\otimes d_{x^{2}-y^{2}} $ interactions towards the total MAE in ABA(ABC) stacked Cr$_{2}$NO$_{2}$.

In Figure \ref{Fig:7} and \ref{Fig:8} (b)-(c), the contributions of various $ d_{i}\otimes d_{j} $ interactions are shown for Cr\textsubscript{2}NF\textsubscript{2}.Since, irrespective of stacking,  the Cr atoms in this compound are inequivalent, their contributions to the magnetic anisotropy energy (MAE) also differ. In the ABA stacked case, the contributions from $ Cr^{I} $ and $ Cr^{II} $ are 0.11 meV and 0.041 meV, respectively. in the ABC stacked case, contributions from $ Cr^{I} $ and $ Cr^{II} $ are -0.11 meV and 0.049 meV, respectively. Thus, we observe that the contribution from $ Cr^{II} $ towards total MAE is nearly the same in either stacking. More importantly, it is positive in both cases. This implies that the differences in the total MAE when stacking changes are solely due to Cr$^{I}$. Although the magnitude-wise contribution from Cr$^{I}$ is the same in both stacking, one is positive, and the other is negative. To understand this in further detail, we look at the orbital resolved contributions (Figures \ref{Fig:7}(b,c) and \ref{Fig:8}(b,c)). In ABA stacking,$  d_{yz}\otimes d_{z^{2}} $ interaction is the dominant contributor to Cr$^{I}$. This is large and positive. Some negative contribution from $ d_{xy}\otimes d_{x^{2}-y^{2}} $ interaction and a small positive contribution from $ d_{yz}\otimes d_{x^{2}-y^{2}} $ interaction add to the total making it positive and large.In ABC stacking, the positive contribution from the $  d_{yz}\otimes d_{z^{2}} $  interactions reduces significanty. The main contribution to MAE now is from $ d_{xy}\otimes d_{x^{2}-y^{2}} $ . This contribution is negative and slightly larger in magnitude than that in ABA stacked compound . This is primarily responsible for the negative MAE in the ABC stacked compound. The contributions from interactions between other $d$-orbitals cannot change the overall qualitative nature. The origin of such a difference lies in the differences in the electronic structure. Comparing $d$-orbital resolved densities of states of  $ Cr^{I} $  in ABA and ABC stacked compound (Figure S18, supplementary information), we observe that in ABC stacked case, $ d_{yz}/d_{z^{2}} $ states in the unoccupied part of the spectrum are closer to Fermi level. Consequently, the coupling between the occupied and unoccupied states of the same spin is enhanced. This enhances the negative contributions and provides more negative contributions to $ d_{yz}\otimes d_{z^{2}} $ in comparison to that in the ABA stacked compound. This also leads to contributions from  $ d_{xy}\otimes d_{x^{2}-y^{2}} $ interaction that is more negative (than in ABA stacked compound). A similar analysis for all other MXenes considered in this work using the orbital-resolved MAE (Figures S24-S25, supplementary information) can explain the qualitative changes in their magnetic anisotropy as stacking changes. 
\section{Conclusions}
MXenes have drawn attention recently due to immense tunability in them, originating from their compositional and structural flexibility. However, so far, the option of flexibility in stacking the transition metal layers hasn't been investigated, primarily because of the fact that any stacking other than the conventional ABC one wasn't discovered experimentally. Taking a cue from the experimental discovery \cite{urbankowski20172d} and subsequent theoretical calculations of several MXenes \cite{gouveia2020mxenes}, we have investigated in detail the effects of stacking patterns on the electronic and magnetic properties of various nitride MXenes functionalized with two different functional groups. We find that stacking adds to the degree of flexibility, leading to a number of different magnetic and ground states in the same composition. We find that transitions between various AFM orders, AFM to FM, and even non-magnetic to FM, are possible due to this.Moreover, the electronic ground states can also be tuned by changing the stacking, leading to transformations like non-magnetic metal to spin gapless semiconductors and metal to half-metal. Such tunability makes these systems suitable for several magnetism-driven applications. Even large magnetic anisotropy energy, comparable to a few well-known 2D magnets and magnetic transition temperatures close to room temperature, are obtained upon changing the stacking pattern from conventional ABC to ABA. These certainly should provoke more investigations in this direction for other systems in the MXene family. We have analyzed the obtained results in great detail and made connections to the structural parameters and the electronic structures of the systems, resulting in a lucid picture of the microscopic physics responsible for the fascinating properties obtained. Although all but one MXene with unconventional stacking has been discovered, the calculations in Reference \onlinecite{gouveia2020mxenes} have clearly shown that in MXenes, the transformation from conventional ABC to unconventional ABA would not cost a large amount of energy. Therefore, even though energetically, ABA stacking may not be preferable in some of the cases, the transformation to it can be easily achieved. This is an essential aspect that can be exploited to utilize the tunability of magnetic MXenes established in this work to further advantage.    

%\bibliographystyle{rsc} % or  "apsrev4-2" "apsrmp4-2" "plain", "unsrt", %"alpha", "abbrv", etc.
%\bibliography{ref}

\providecommand*{\mcitethebibliography}{\thebibliography}
\csname @ifundefined\endcsname{endmcitethebibliography}
{\let\endmcitethebibliography\endthebibliography}{}

\end{document}